\documentclass[aip,jcp]{revtex4-1}
\usepackage{epsfig}

\begin{document}

\title{A simple protocol for the probability weights of the simulated 
tempering algorithm: applications to first-order phase transitions}
\author{Carlos E. Fiore} 
\email{fiore@fisica.ufpr.br}
\author{M. G. E. da Luz}
\email{luz@fisica.ufpr.br}
\affiliation{Departamento de F\'{\i}sica, 
Universidade Federal do Paran\'a, 
CP 19044, 81531-990 Curitiba, Brazil}  
\date{\today}

\begin{abstract}
The simulated tempering (ST) is an important method to deal with 
systems whose phase spaces are hard to sample ergodically.
However, it uses accepting probabilities weights which often demand 
involving and time consuming calculations. 
Here it is shown that such weights are quite accurately obtained 
from the largest eigenvalue of the transfer matrix -- a quantity  
straightforward to compute from direct Monte Carlo simulations -- 
thus simplifying the algorithm implementation. 
As tests, different systems are considered, namely, Ising, Blume-Capel, 
Blume-Emery-Griffiths and Bell-Lavis liquid water models.
In particular, we address first-order phase transition at low 
temperatures, a regime notoriously difficulty to simulate 
because the large free-energy barriers.
The good results found (when compared with other well established 
approaches) suggest that the ST can be a valuable tool to address
strong first-order phase transitions, a possibility still not well 
explored in the literature.
\end{abstract} 

\pacs{05.10.Ln, 05.70.Fh, 05.50.+q}  
\keywords{simulated tempering, first-order phase transitions, 
eigenvalues of the transfer matrix, Ising and lattice-gas models} 

\maketitle   


\section{Introduction}

Many statistical systems are difficult to ``probe'' since their 
phase spaces display complicated landscapes full of energetic 
valleys and hills \cite{mauro1,helmut}.
In such case, a non-representative sampling of the microstates, 
e.g., due to uneven visits to the different domains \cite{williams},
can lead to metastability and broken ergodicity, with a consequent
non-convergence to equilibrium and poor estimates for the thermodynamic 
quantities \cite{palmer,neirotti}.
This is exactly the situation found in first-order phase transitions 
\cite{besold}, where the free-energy minima are separated by large 
barriers, and simple one-flip Metropolis approaches are unable to 
solve the problem.

Thus, different methods -- aimed to guarantee ergodic simulations 
in the context described above -- have been proposed 
\cite{cluster1,rachadi,cluster2,stolovitzky,wang,earl,gao}.
Among the so called enhanced sampling algorithms, a particularly 
important one due to its simplicity and generality is the simulated 
tempering (ST) \cite{marinari,pande} (closely related to the also
relevant parallel (or replica) tempering, PT, approach \cite{pt}).
Here, tempering means that along the simulations the system can undergo 
temperature changes.
Consider we shall analyze a system at $T = T_1$.
Besides an usual Monte Carlo (MC) prescription, in the ST algorithm 
$T$ is also treated as a dynamical variable:
the temperature can assume distinct values from a set 
$\{T_1 < T_2 < \ldots < T_N\}$, switching from time to time according 
to established accepting probabilities $p$'s. 
Of course, during the simulations the relevant averages necessary 
to obtain the sought thermodynamic quantities are performed only
when the system is at $T_1$.
The central idea is that temporary evolutions at higher $T$'s 
strongly facilitates the crossing of the free-energy barriers, then 
allowing uniform visits to the multiple regions of a fragmented 
phase-space \cite{berg}.

A fundamental ingredient in the ST is the definition of the accepting 
probabilities \cite{marinari}; functions of the energy, the different 
$T$'s and appropriate weight factors $g$ (see next Section as well as
interesting discussions in Ref. \cite{kim-heinsalu-fasnacht}).
By their turn, the exact expressions for the $g$'s depend on the 
problem partition function $Z$, a quantity often difficult to calculate 
\cite{sauerwein,landau-binder,ma}, demanding computationally time 
consuming methods \cite{okamoto}.
Obviously, approximations for the $g$'s can be used (e.g., as in Ref. 
\cite{pande}).
But so the ST efficiency can be dramatically hindered, once the 
weights themselves can create certain bias for the sampling.
For instance, they may lead to unbalanced sorted temperatures, given
rise to a non-ergodic visit of the phase-space if the probabilities 
to pick the lower $T_n$'s are too high.
On the other hand, if the more frequent $T$'s are the greater ones, 
the generated set of microstates cannot properly characterize the system 
features at $T = T_1$, the actual temperature of interest.
Such technical aspects associated to the $g$'s are even more delicate 
for first order phase transitions, a situation rarely analyzed with 
the ST \cite{besold,st-first-order}.

Given the previous comments, our purpose in this contribution is
twofold.
First, to present a numerically simple -- yet quite accurate
-- procedure to obtain the weight factors $g$, thus making the ST
algorithm easier to implement, e.g., by avoiding involving recursive 
protocols \cite{ma}.
This is accomplished with a method proposed in Ref. \cite{sauerwein}, 
where $Z$ is calculated directly from the transfer matrix 
largest eigenvalue ($\lambda^{(0)}$), straightforward to compute from 
Monte Carlo simulations.
The key point is that although $\lambda^{(0)}$ gives the exact $Z$ 
only at the thermodynamical limit, i.e., for infinite systems, the 
convergence is very fast.
So, even for a relatively small system (as will be illustrated 
in the examples), any difference between its exact $Z$ and that 
from $\lambda^{(0)}$ can be neglected and for all practical reasons 
the approach leads to the correct $g$'s.
Second, to consider the ST for the already mentioned difficult case 
of first-order transitions, showing that the ST is also a helpful 
tool to address such regime, a possibility barely explored in the 
literature.

The work is organized as the following.
We review the ST algorithm and how to calculate $g$ from the transfer 
matrix in Section II.
Also in Section II we exemplify the procedure efficiency by computing 
the partition function for the Ising model, a system for which $Z$ 
can be obtained exactly.
In Section III and IV we compare the present with other well
established methods to study first-order phase transitions, taking 
as case studies the Blume-Capel, Blume-Emery-Griffiths (BEG) and 
Bell-Lavis models.
Finally, remarks and the conclusion are drawn in Section V.

\section{The ST and the calculation of $g$}

As mentioned in the Introduction, the ST algorithm is generally 
implemented as two steps procedure, repeated a given number of 
times.
First, at a temperature $T_{n'}$ ($n' = 1, 2, \ldots, N$), a 
standard Metropolis prescription is used to promote the transition 
$\sigma' \rightarrow \sigma''$ with the probability
$P_{\sigma' \rightarrow \sigma''} = 
\mbox{min}\{1, \exp[-\beta_{n'} 
({\mathcal H}(\sigma'') - {\mathcal H}(\sigma'))]\}$ 
(for $\sigma$ representing the system microscopic configurations).
Second, an attempt to change the replica temperature
(from $T_{n'}$ to $T_{n''}$) is made according to 
\begin{equation}
p_{n' \rightarrow n''} = 
\min \{ 1, \, \exp[(\beta_{n'} - \beta_{n''}){\mathcal H}(\sigma) 
+ (g_{n''} - g_{n'})] \}.
\label{p-definition}
\end{equation}
In Eq. (\ref{p-definition}), ${\mathcal H}$ is the problem 
Hamiltonian, $\sigma$ the actual microscopic state, and 
$n'' = 1, \ldots, N$ arbitrary, 
so non-adjacent temperatures changes are allowed.
Moreover, $\beta = 1/(k_B \, T)$ with $k_B$ always set equal to 1
hereafter.

We observe that $p_{n' \rightarrow n''}$ is strongly dependent on the 
weights $g$'s.
Indeed, for an appropriate sampling the evolution should uniformly 
visit all the established temperatures, which is the case when
$g_{n} = \beta_{n} \, f_{n}$ for $f_n$ the free-energy per volume 
$V$ at $T_n$.
Recalling the relation $\beta_n \, f_n = - \ln[Z_n]/V$, we see that 
the calculation of $g$ is not a trivial task:
neither the partition function nor the free-energy can be obtained 
directly from MC simulations since there are no thermodynamic 
quantities whose averages lead to $Z$ and $f$.
For this reason some alternative methods have been proposed 
\cite{pande,park2,pande2}.
Here we shall consider a rather simple numerical approach to compute 
$f_n$ \cite{sauerwein}, based on the transfer matrix ${\mathcal T}$ 
largest eigenvalue $\lambda^{(0)}$.

Briefly, at the thermodynamic limit it holds true that 
(see details in the Appendix)
\begin{equation}
Z_n = (\lambda^{(0)}_n)^K,
\label{z-lambda}
\end{equation} 
with $K \sim V^{1/d}$ and $d$ the spatial dimension.
To obtain $\lambda^{(0)}$ is straightforward.
In fact, suppose for definiteness a 2D system with $K$ layers 
of $L$ sites each, so $V = L \times K$.
Next, consider the full Hamiltonian decomposed as
\begin{equation}                                                               
{\cal H}= \sum_{k=1}^{k=K} {\cal H}(S_k, S_{k+1}),                                
\label{e14}                                                                    
\end{equation}
where $S_k \equiv (\sigma_{1,k}, \sigma_{2,k}, \ldots, \sigma_{L,k})$
represents the state configuration of the $k$-th layer.
We further assume periodic boundary conditions, or $S_{K+1} = S_1$.
The transfer matrix is defined (Appendix) so that its elements 
read
\cite{sauerwein}
\begin{equation}
{\mathcal T}(S_{k}, S_{k+1}) =  
\exp[- \beta \, {\cal H}(S_k, S_{k+1})].
\end{equation} 
Then, as shown in the Appendix, we have
(${\mathcal T}(S_k, S_{k+1} = S_k) \equiv {\mathcal T}(S_k))$
\begin{equation}
\lambda^{(0)} = \langle {\mathcal T}(S_k) \rangle / 
\langle \delta_{S_{k}, S_{k+1}} \rangle. 
\label{lambzero}
\end{equation}                                                
This expression enables one to calculate the largest eigenvalue 
$\lambda^{(0)}$ of ${\mathcal T}$ in terms of the 
averages $\langle {\mathcal T} (S_k) \rangle$ and 
$\langle \delta_{S_k, S_{k+1}} \rangle$, with 
$\delta_{S_k, S_{k+1}} = 1$ ($\delta_{S_k, S_{k+1}} = 0$)
if the layers $S_k$ and $S_{k+1}$ are equal (different). 
We also should mention that $\langle {\mathcal T} (S_k) \rangle$ and 
$\langle \delta_{S_k, S_{k+1}} \rangle$ can be evaluated 
from quite standard MC simulations.

As the final step, the weights follow from
\begin{equation}
g = -\ln[\lambda^{(0)}_n]/L.
\label{g-eigen}
\end{equation}
A relevant issue is that even thought Eq. (\ref{z-lambda}) is exactly 
only for infinite size systems, if $L$ (or $K$) is not too small the 
relation is extremely accurate.
Hence, for any practical purpose Eq. (\ref{g-eigen}) gives the 
correct $g$, as we are going to illustrate in the next Sections.

In this way, we can summarize the proposed approach as the following.
First, with Eqs. (\ref{lambzero})-(\ref{g-eigen}) one evaluates the 
partition function and consequently the free-energy weights (at 
the temperatures $\{T_n\}$) from usual MC simulations. 
Second, having the correct $g$'s, one just implement the ST 
algorithm as previously explained.

Finally, we comment on three important technical issues.
The first is related to the query: how can standard MC simulations 
lead to good values for $g$ around first-order phase transition if 
then Metropolis algorithm usually yields unbalanced samplings?
The answer relies on the fact that the free-energies are all equal
at such regime of phases coexistence.
Hence, even a biased microscopic sampling will result in accurate 
free-energies and consequently appropriate $g$'s.
On the other hand, the metastability typical of $T$'s in the vicinity
of the transition temperature can difficult the determination 
(with the necessary precision) of the function $Z \times T$ in such 
interval.
Thus, derivatives of $Z$ with respect to distinct parameters,
representing different thermodynamic quantities, will give poor results.
So, a second point is that the transfer matrix alone is not
a reliable method to study first-order phase transitions.
Lastly, we mention that in the limit of large systems and high
temperatures $\langle \delta_{S_k, S_{k+1}} \rangle$ is small, since the 
probability for the configurations $S_k$ and $S_{k+1}$ to be the same 
is very low.
As a consequence, one may get bad estimations for the weights. 
In this case, a possible way to circumvent the problem is to 
decrease the size $L$ of each layer and to increase the number of 
layers, maintaining the volume $V = L \times K$ constant (see also
the discussion in the Section V).

\subsection{An example: the partition function for the Ising model}

Just to verify how good is the Eq. (\ref{z-lambda}) for finite 
size systems, we consider the Ising model, whose partition function 
can be calculated exactly.
The Hamiltonian is 
\begin{equation}                                                            
{\cal H} = - J \sum_{<i,j>} \sigma_{i} \, \sigma_{j} 
- H \sum_i \sigma_i,       
\end{equation}
where $<i,j>$ denotes nearest-neighbors pairs $i$ and $j$ in a 
lattice of $V = L^{d}$ sites. 
At each site $i$, the spin variable assumes the values $\sigma_i = \pm 1$.
$J$ is the interaction energy and $H$ is the magnetic field. 
For a square lattice ($d=2$), ${\mathcal T}(S_{k})$ yields
\begin{equation}                                                                
{\mathcal T}(S_{k}) = \exp\left[ \beta \, \big(
\sum_{l=1}^{L} J \,
(1 + \sigma_{l,k} \, \sigma_{l+1,k}) + H \, \sigma_{l,k} \big) \right].      
\label{e17}                                                                     
\end{equation}

%
\begin{figure}
\setlength{\unitlength}{1.0cm}
\includegraphics[scale=0.4]{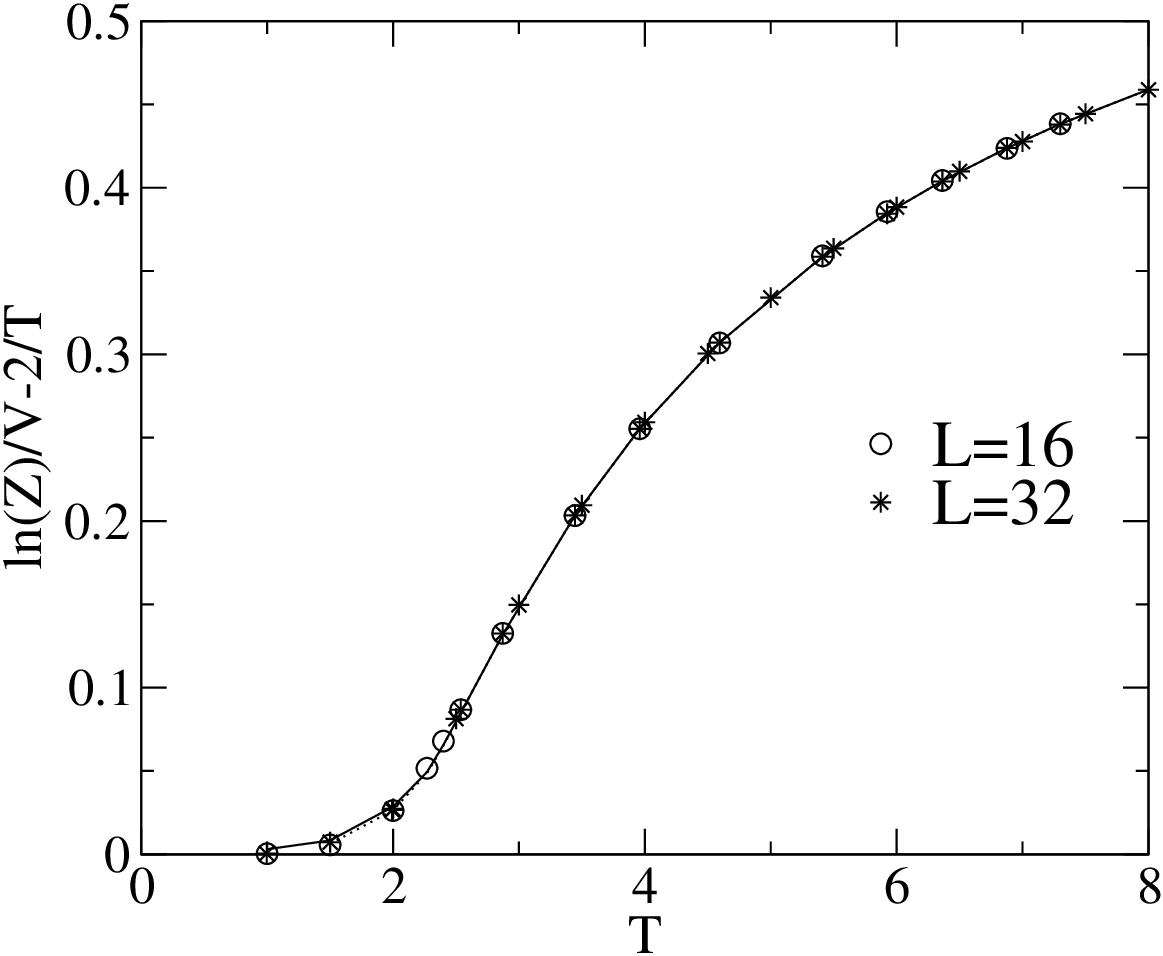}
\caption{For the Ising model, the partition function versus 
$T$ calculated exactly \cite{ferdinand} and from 
Eq. (\ref{z-lambda}) (symbols), with the parameters as in the text.
The exact solutions, the lines, for $L=16$ (continuous) and 
$L=32$ (dotted) are almost indistinguishable.}
\label{partising}
\end{figure}
%

Numerical calculations have been performed for $H=0$ (and in 
units of $J$).
For $L=16$ and $L=32$, Fig. \ref{partising} compares the 
exact partition function for a finite system (obtained from the 
solution in Ref. \cite{ferdinand}) with that calculated from Eq. 
(\ref{z-lambda}).
The agreement is indeed remarkable, indicating that even for 
relatively small systems, $Z$ and therefore $f$ are already 
very close to their values at the thermodynamic limit.

Since the above system presents a very well known and simple
first-order phase transition (for $H=0$ and $T < T_c$), we prefer 
to address such regime, our focus in this contribution, for other 
models in the following Sections.

\section{The lattice-gas model with vacancies (BEG)}

The lattice-gas with vacancies (BEG) model is given by the 
Hamiltonian ($\sigma = 0, \pm 1$)
\begin{equation}                           
{\cal H} = 
-\sum_{<i,j>} (J  \, \sigma_{i} \, \sigma_{j} + K \, 
\sigma_{i}^{2} \, \sigma_{j}^{2}) 
- \sum_{i} (H \, \sigma_i - D \, \sigma_i^2),
\label{e3}                                            
\end{equation}
for which the transfer matrix $\mathcal{T}(S_k)$ elements are 
\begin{equation}                                                
{\mathcal T}(S_{k}) = 
\exp\Big[ \beta \sum_{l=1}^{L} 
\Big( (H + J \, \sigma_{l+1,k}) \, \sigma_{l,k} 
+ (J - D + K \, (1 + \sigma_{l+1,k}^{2})) \, \sigma_{l,k}^{2} 
\Big) \Big].
\label{e18}
\end{equation}

For the values of $K/J$ we are going to consider here and at low
temperatures, if $D$ is small, the system presents an ordered
phase.
When $D$ increases, a gas phase takes place.
These regimes are separated by a strong first-order phase transition
at $D = D^{*}$.
For simplicity, hereafter all the quantities will be given in
units of $J$.

\subsection{The Blume-Capel model}

\begin{figure}
\setlength{\unitlength}{1.0cm}
\includegraphics[scale=0.55]{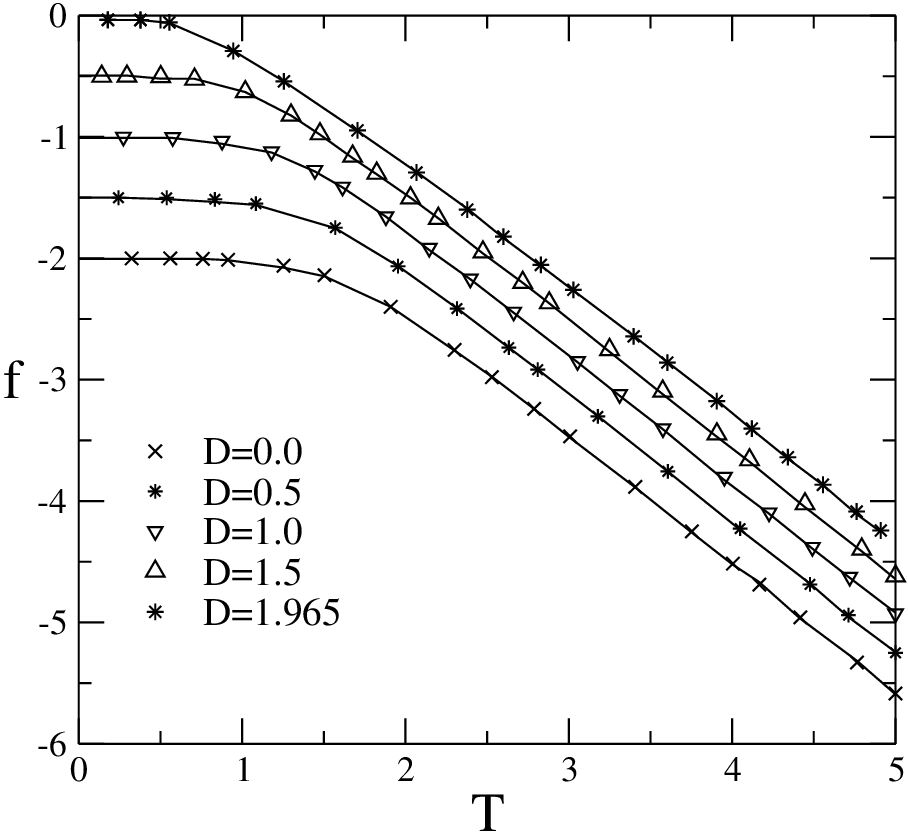}    
\caption{For the Blume-Capel model, the free-energy versus $T$ 
for different $D$ obtained from the present approach (symbols)
and from the Wang-Landau method (continuous lines).}
\label{compblume}
\end{figure}

For the Blume-Capel case, $K=0$, accurate estimates for $f$
(which directly gives the ST weights once $g = \beta \, f$) is 
available from the very efficient Wang-Landau method.
We then compare in Fig. \ref{compblume} the free-energy from our 
proposed procedure with that from Wang-Landau's \cite{pla}, 
considering $K=0$, $L=32$ and different $D$'s.
As for the Ising model, again we see an excellent agreement (even 
for $D = 1.965$, the value corresponding to the tricritical point 
for $T$ about 0.609).
We also have explicit tested smaller $L$'s (down to 12), always
obtaining very good results.
So, the exact $g$'s for the ST are adequately (and easily) obtained 
from the transfer matrix approach.

\begin{figure}
\setlength{\unitlength}{1.0cm}
\includegraphics[scale=0.34]{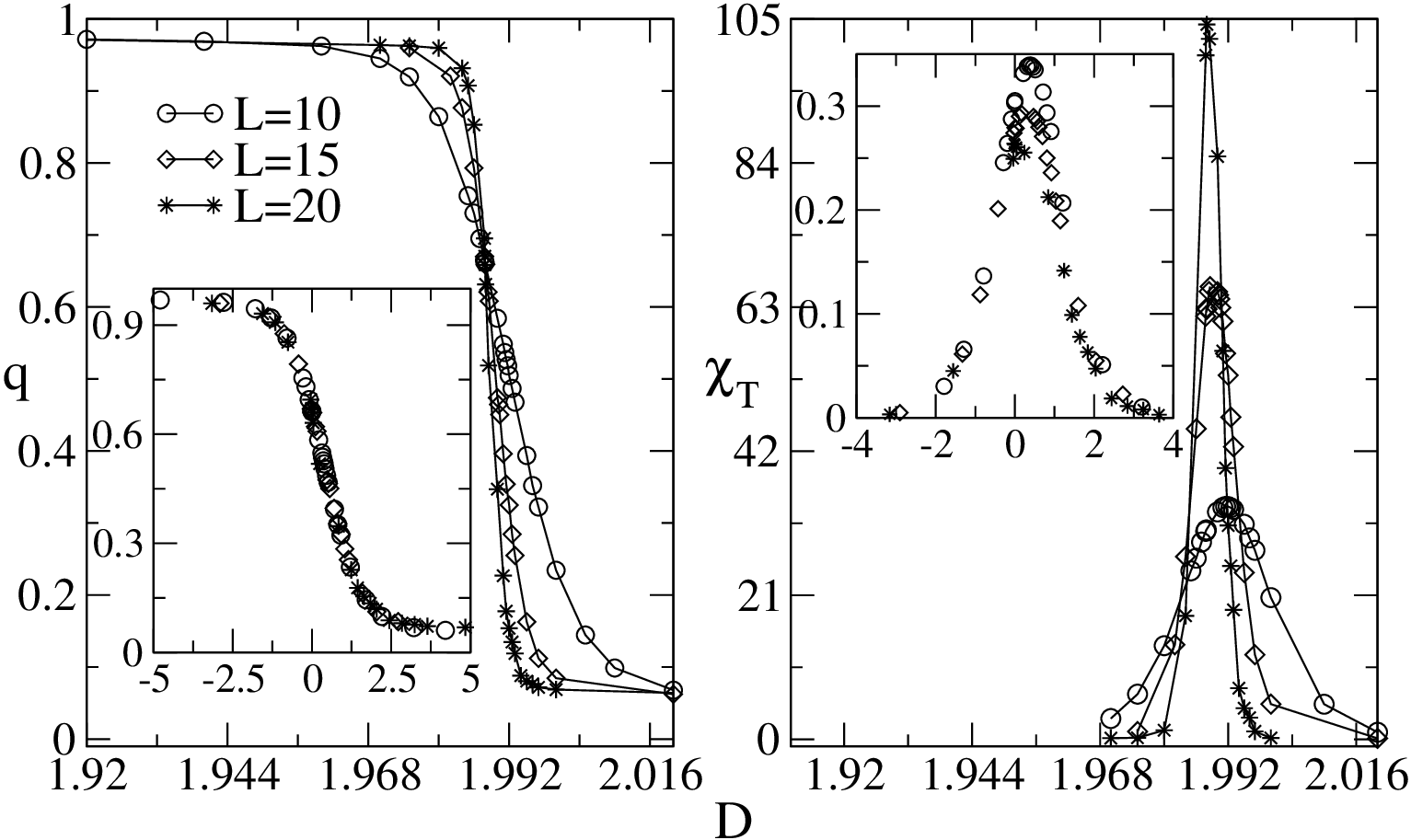}    
\caption{For the Blume-Capel model, the order parameter $q$ and 
the isothermal susceptibility $\chi_T$ versus $D$ for different 
system sizes at $T=0.5$. 
The insets show the collapsed data, respectively, plotted as
$q \times (D - D^{*}) \, V$ and 
$\chi_T/V \times (D - D^{*}) \, V$.}
\label{blumefs}
\end{figure}
%
%

To study the Blume-Capel model around first-order phase transition,
we perform finite size scaling analysis using the PT algorithm.
We first recall that according to a rigorous theory of first-order 
phase transitions at low temperatures \cite{borgs}, all thermodynamic 
quantities should scale with $V$.
Then, for $Q = \sum_{i=1}^{V} \sigma_i^2$, in Fig. \ref{blumefs} 
we show the order parameter 
$q = \langle Q \rangle/V$ and the isothermal susceptibility
$\chi_T = (k_B \, T)^{-1} (\langle Q^2 \rangle - \langle Q \rangle^2)/V$ 
as functions of $D$ for three system sizes $L$.
The transition point can be estimated, for instance, from the peak 
position in the susceptibility or in the specific heat curves. 
Here we obtain $D^{*}$ through the location of the distinct 
$L$ isotherms crossing \cite{fiore4,cluster2,fiore8,fernades-levin}.
We find that for $T=0.50$ all the isotherms cross at 
$D^{*}=1.9879(1)$, which closely agree with the values 
$T=0.499(3)$ and $D^{*}=1.992(1)$ in Ref. \cite{pla}.
In addition, the good collapse of the data in the insets,
plotted as $q \times (D-D^{*}) \, V$ and as
$\chi_T/V \times (D-D^{*}) \, V$, 
illustrates the adequacy of the ST to locate transition points.
Of course, other exact implementations of the ST, like that in 
Ref. \cite{ma}, would also solve the problem, however, by means of
more complicated protocols, thus demanding longer computational
times.

\subsection{The full, $K \neq 0$, BEG model}

Next, we briefly comment on the full BEG model, i.e., $K \neq 0$.
Generally, approaches based on cluster algorithms are known to be
very appropriate for lattice-gas systems \cite{cluster1}.
Nevertheless, it is also a fact that first-order phase transitions 
for some special $K$'s, like $K=3.3$, can be a little trick to 
solve.
So, modifications in the cluster method are necessary \cite{rachadi}.

\begin{figure}
\setlength{\unitlength}{1.0cm}
\includegraphics[scale=0.37]{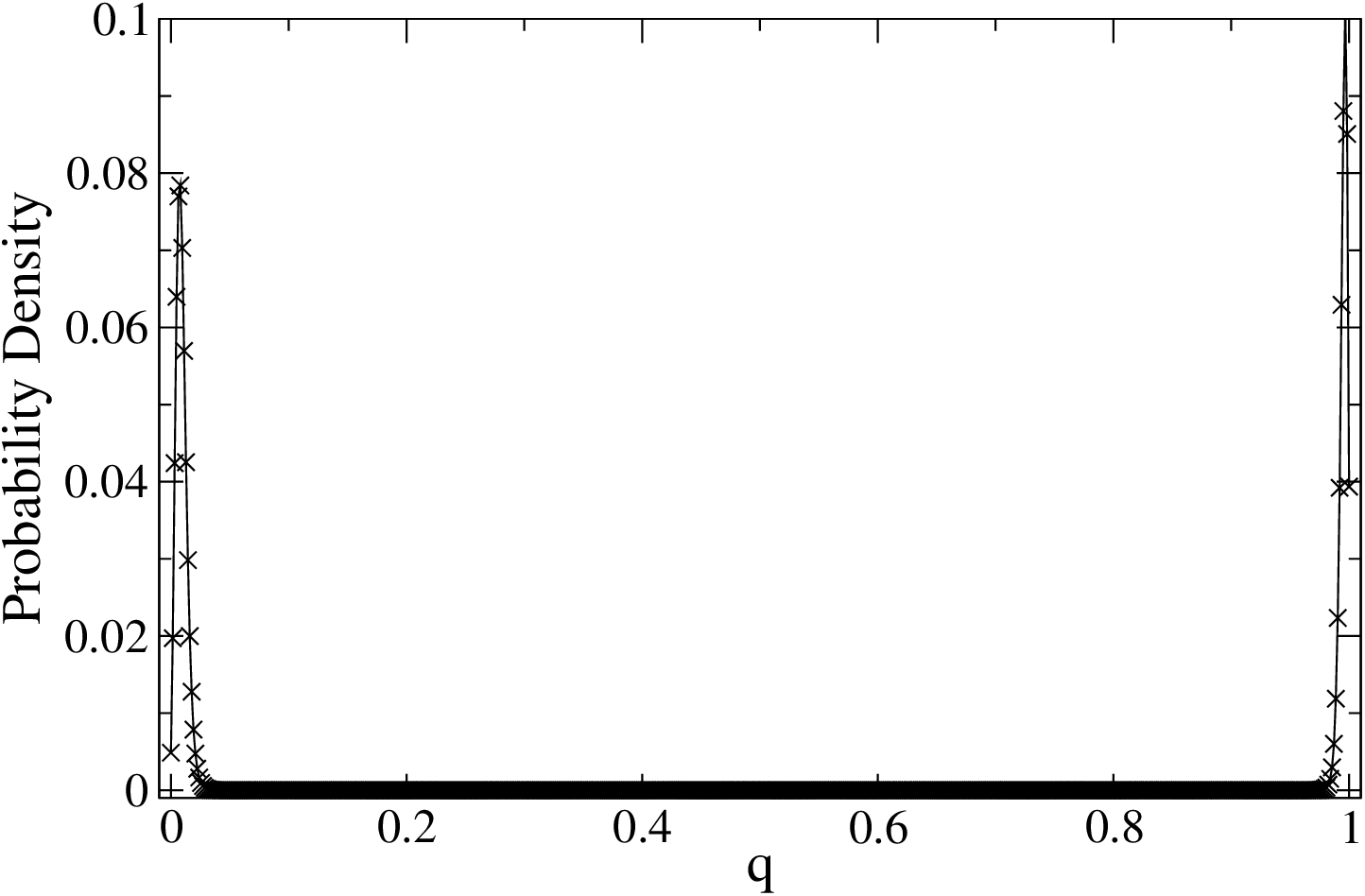}
\caption{For the BEG model, with $K/J = 3.3$ and parameters
as in the text, the probability distribution histogram of the order 
parameter $q$. The continuous line is a guide for the eyes.}
\label{hist-st-beg}
\end{figure}

In the absence of calculations of $f$ and $Z$ when $K=3.3$
in the literature (for a explicit comparison), in Fig. \ref{hist-st-beg} 
we just present the probability distribution histogram for the 
order parameter $q$, considering $D = 8.605$, $T=1.50$ and a small 
lattice of $L=25$.
From the two peaks with very similar heights, we see that the 
simulations are able to cross the large free-energy barriers at the 
phase coexistence.
On the other hand, from an usual Metropolis simulations, the system 
would be trapped in metastable states, evolving to the stable phases 
only after very large MC steps.
We should mention that qualitatively our results agree quite well 
with those in Fig. 1 (b) of Ref. \cite{rachadi} for very similar
parameters (actually, there the authors use a bigger lattice of 
$L=32$ at $T=1.50$, with $D = 8.6035$).
We do not make a direct quantitative comparison, e.g., by 
digitalizing the results in Ref. \cite{rachadi}, simple because in 
the mentioned figure the scales are not explicitly given.
By setting $L=32$ (for which the eigenvalue method is even 
better because the greater $L$), we also have found a balanced bimodal 
distribution at the coexistence, with $D \approx 8.6035$ as
in Ref. \cite{rachadi}.

As a final observation, we recall that $K=3.0$ has been extensively 
studied under different approaches \cite{cluster2,fiore8,fiore4}.
In particular, it has been shown \cite{fiore-luz} that for this
parameter value, the ST is an efficient method to study the 
first-order phase transition for 2D square lattices as small as 
$L=20$.

\section{The Bell-Lavis model}

As a last example, we consider the Bell-Lavis water model 
\cite{bell,bell2}. 
It is defined on a triangular lattice, whose occupational variable
$\sigma_i$ ($i=1, 2, \ldots, V$) takes the values 1 (0) if the 
site is (is not) occupied by a water molecule. 
Moreover, each molecule is described by an orientational state,
indicating to which nearest neighbor a hydrogen bonding can be 
formed.
So, for a non-empty site $i$, we define the quantity $\tau_{i}^{ij}$, 
where $j$ runs over the first neighbors $j$ of $i$.
If there is a bonding arm pointed from $i$ towards $j$, then 
$\tau_{i}^{ij} = 1$, otherwise $\tau_{i}^{ij} = 0$.
Two adjacent molecules always interact via Van der Waals forces, 
whereas they do form hydrogen bonds provided 
$\tau_{i}^{ij} \times \tau_{j}^{ji} = 1$.
 
The model is described, in the grand-canonical ensemble, by the
Hamiltonian
\begin{equation}                                         
{\mathcal H} = -\sum_{<i,j>} \sigma_{i} \, \sigma_{j} \,
(\epsilon_{hb} \, \tau_{i}^{ij} \, \tau_{j}^{ji} + 
\epsilon_{vdw}) - \mu \sum_{i} \sigma_{i},                                        
\label{hambl}                                                                
\end{equation}
where $\mu$ is the chemical potential and $\epsilon_{vdw}$ and 
$\epsilon_{hb}$ are, respectively, the Van der Waals and hydrogen 
bonds interaction energies.
The Van der Waals force tends to increase the system density by 
filling the lattice with molecules.
On the other hand, the hydrogen bond interaction essentially favors 
an increasing in the hydrogen bonds, so it effectively may limit 
the molecule density if $\mu$ is negative and small.
Finally, the ${\mathcal T}(S_k)$ elements are given by
\begin{equation}
T(S_{k}) = \exp\Big[\sum_{i=1}^{L}\Big(
\sigma_{i,k} \, (\sigma_{i,k} + 2 \sigma_{i+1,k}) 
(\epsilon_{vdw} + \epsilon_{hb} \, \tau_{i,k} \,
\tau_{i+1,k} + \mu) \Big) \Big].
\end{equation}

The Hamiltonian in Eq. (\ref{hambl}) exhibits a very rich phase 
diagram. 
For instance, recent numerical simulations \cite{fiore11} show 
that the parameter conditions allowing the existence of two stable
liquid phases take place when the hydrogen bonds are at least 
three times higher than the Van der Wall interaction, or 
$\zeta \equiv \epsilon_{vdw}/\epsilon_{hb} < 1/3$. 
In such case, for low negative $\mu$ values the system presents 
only a gas phase. 
By increasing $\mu$ a low-density-liquid-phase (LDLP) arises.
In the limit of higher chemical potentials, we have a 
high-density-liquid-phase (HDLP).
At $T = 0$, the LDLP (HDLP) has a global density of 
$\rho=2/3$ ($\rho=1$), with the hydrogen bond density per molecule 
being $\rho_{hb}=3/2$ ($\rho_{hb}=1$). 
Furthermore \cite{fiore11}, both phase transitions are of 
first-order: that between the gas and the $LDLP$ occurring at 
$\mu^{*} = -3 \, (1+\zeta)/2$ and that between the LDLP and 
HDLP at $\mu^{*} = -6 \, \zeta$.
For $T > 0$, the former remains first-order, ending at a 
tricritical point, whereas the latter becomes second-order,
belonging to the Ising universality class \cite{fiore11}.

\begin{figure}
\setlength{\unitlength}{1.0cm}
\includegraphics[scale=0.45]{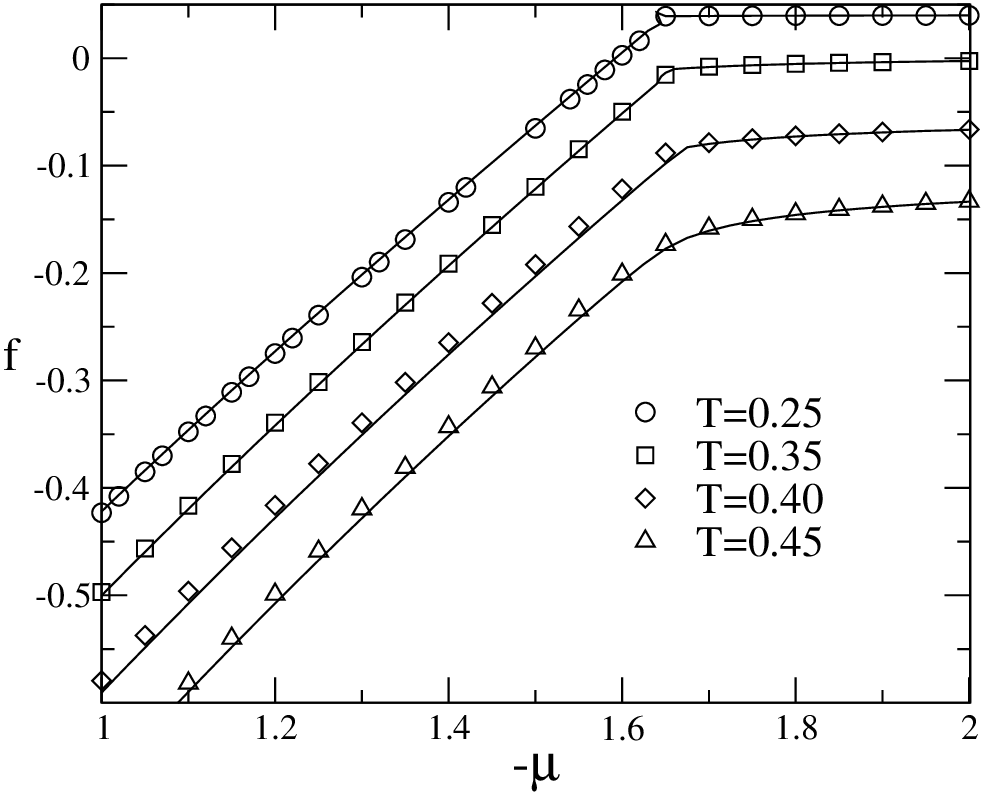}
\caption{For the Bell-Lavis model with $L=18$, $f \times \mu$ 
obtained from the integration of the Gibbs-Duhem relation 
(continuous lines) and from the largest eigenvalue of 
${\mathcal T}$ (symbols). The different curves have been
offset for a better visualization.}
\label{compbell}
\end{figure}

Next, we will focus on the first-order phase transition case, 
gas--LDLP, assuming $\zeta=1/10$ and presenting all the results 
in units of $\epsilon_{hb}$.
First, we test the efficiency of the transfer matrix to yield 
the free-energies, hence the weights for the ST.
In Fig. \ref{compbell} we compare some values of $f$ calculated 
from the ${\mathcal T}$ approach with those obtained from numerical 
integration of the Gibbs-Duhem equation, 
$S \, dT- V \, dp + N \, d\mu = 0$, at fixed temperatures.
Thus, $dp =\rho \, d\mu$ where the pressure is related to the 
free-energy per volume (grand-canonical) by the expression $f=-p$.
As it can be seen, even for a small lattice size of $L=18$ and  
low $T$'s, the agreement is very good.

\begin{figure}
\setlength{\unitlength}{1.0cm}
\includegraphics[scale=0.35]{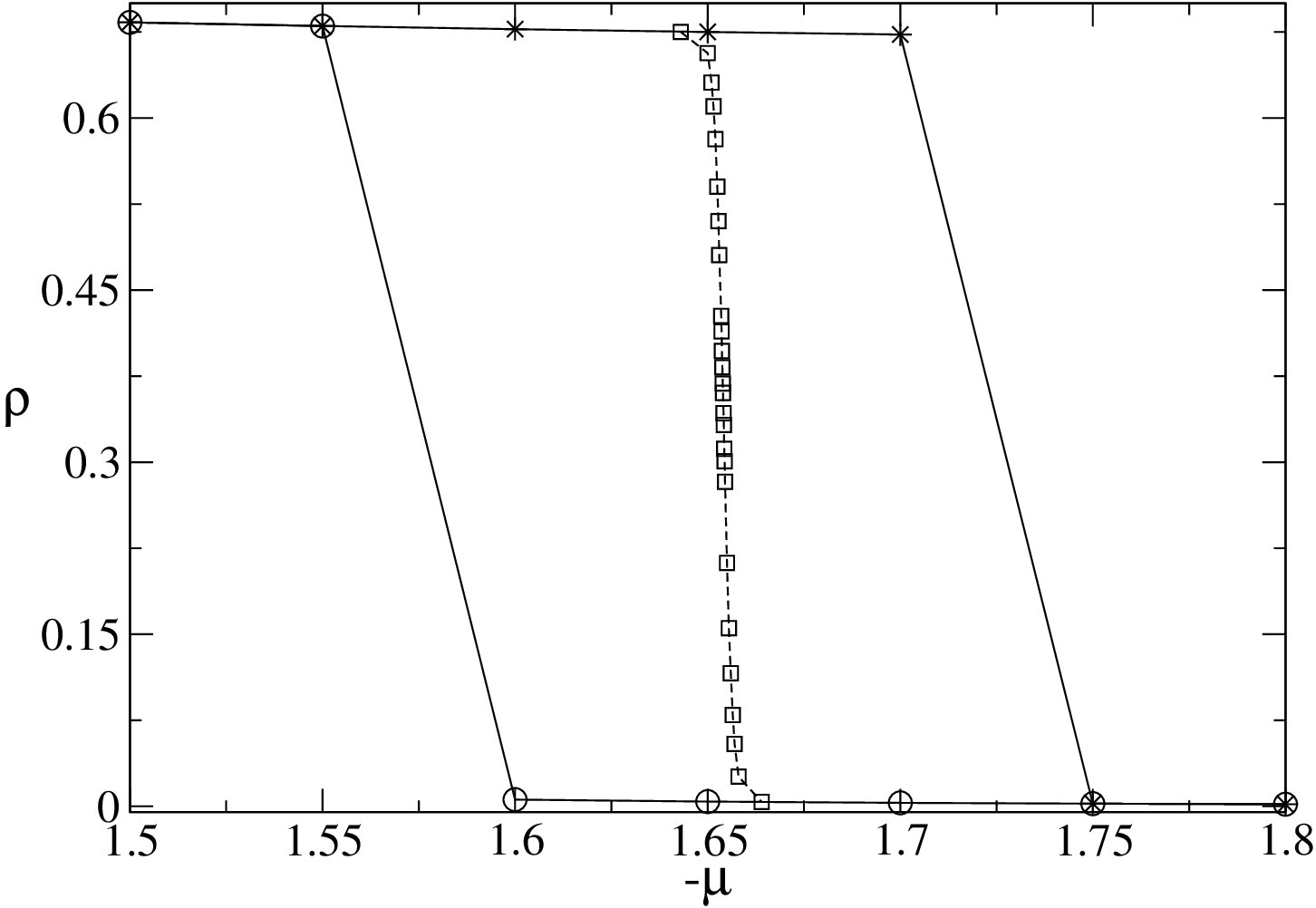}
\caption{For the Bell-Lavis model with $L=24$ and at $T=0.25$, 
$q$ versus $\mu$ calculated from the ST (square) and from a common
Metropolis method.
The hysteresis is due to the latter algorithm difficulty in properly 
sampling the system.} 
\label{bell-lavis-metropolis}
\end{figure}

\begin{figure}
\setlength{\unitlength}{1.0cm}
\includegraphics[scale=0.42]{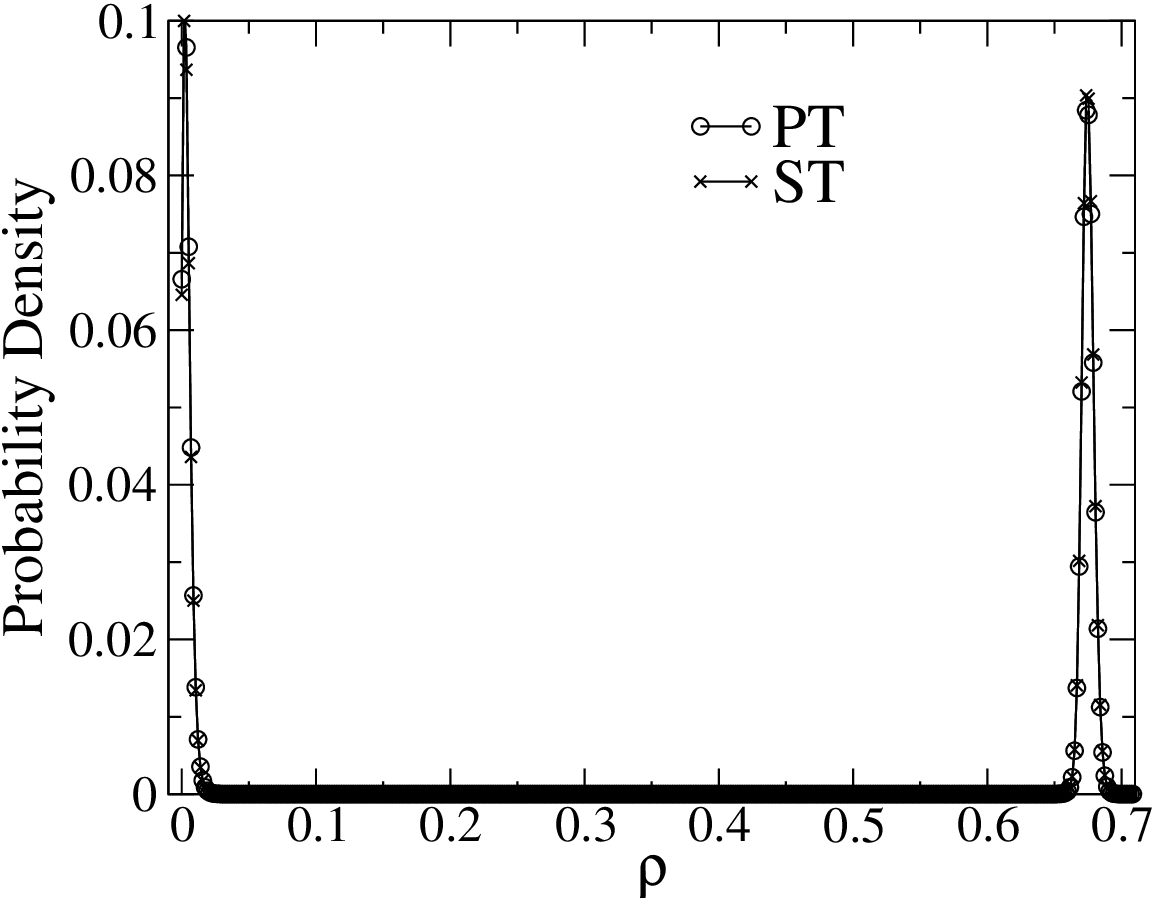}
\caption{For the Bell-Lavis model with $T=0.25$, $L=24$ and  
$\mu=-1.6533$, the probability distribution histogram of the 
order parameter $\rho$  calculated from the ST and PT methods. 
The continuous line is a guide for the eyes.}
\label{hist-pt-st}
\end{figure}

Now, we assume a lattice of $L=24$ and a rather small $T = 0.25$, 
difficult to simulate by standard one-flip algorithms.
In Fig. \ref{bell-lavis-metropolis} we plot the density $\rho$
(the order parameter) as function of the chemical potential 
$\mu$ (the control parameter) around the phase transition
point.
We consider both the ST as well as an usual Metropolis algorithm.
Contrary to the latter, the ST predicts the phase transition without 
displaying any hysteresis.
In fact, hysteresis is a characteristic behavior of methods not able 
to properly sample the system when the phase space presents high 
free-energy barriers.
Also, for the low $T$ considered, the transition $\mu^{*}$ should 
not be too different from -1.65, the thermodynamic value at $T = 0$ 
when $\zeta = 1/10$.
Indeed, an expectation confirmed by Fig. 6.

Due to the lack of studies discussing first-order phase transition 
for the Bell-Lavis model (either by means of general or dedicated 
methods), here we compare the proposed ST algorithm with 
calculations based on the parallel tempering (PT) approach, recently 
shown to be a very efficient tool to analyze such thermodynamical
regime \cite{fiore8,pt-first-order,hansmann2, fiore-luz}.
In Fig. \ref{hist-pt-st} we plot the histogram of the order 
parameter $\rho$ at the phase coexistence for $T=0.25$ and $L=24$.
In `tuning' the chemical potential so to have the two peaks of about
the same high, we find $\mu=-1.6533$.
We emphasize the quite good agreement between the methods, both 
being able to circumvent metastable states and to promote frequent
visits between the gas phase and the LDLP.

\begin{figure}
\setlength{\unitlength}{1.0cm}
\includegraphics[scale=0.34]{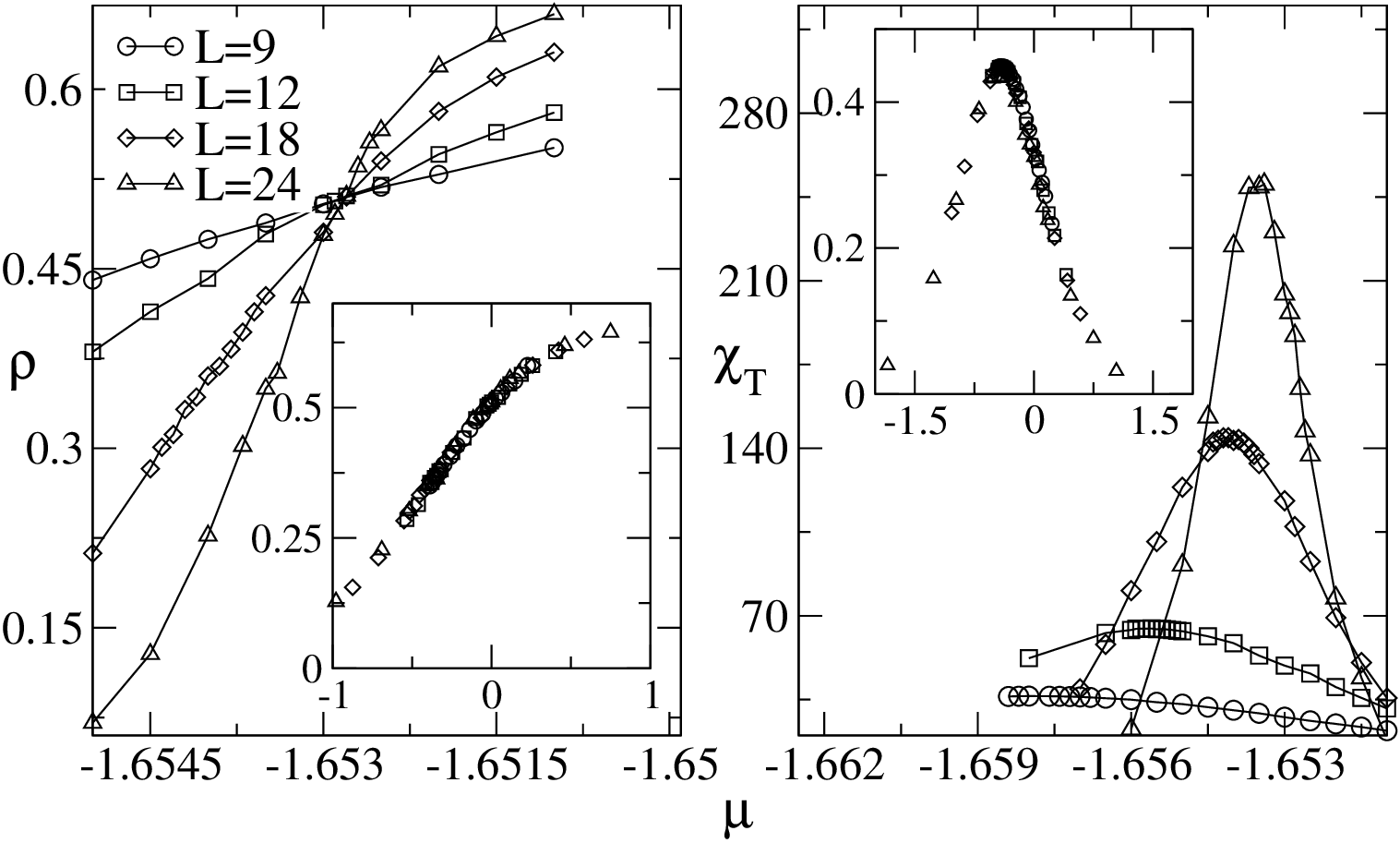}
\caption{For the Bell-Lavis model, the order parameter $\rho$ and 
compressibility $\chi_T$ versus $\mu$ for different 
system sizes at $T=0.25$. 
The insets show the collapsed data, respectively, plotted as
$\rho \times (\mu-\mu^{*}) \, V$ and 
$\chi_T/V \times (\mu-\mu^{*}) \, V$.}
\label{bell-lavis-finite}
\end{figure}

We finally perform a finite-size analysis to determine the phase 
coexistence.
We plot in Fig. \ref{bell-lavis-finite} the order parameter 
$\rho$ and the compressibility $\chi_T$ as functions of $\mu$, 
assuming different system sizes $L$ and fixed $T=0.25$. 
Note that we obtain very smooth curves and that $\chi_T$ 
exhibits sharper peaks as $L$ increases.
Moreover, all the isotherms for the density $\rho$ cross
each other at the same point $\mu^{*}=-1.6528(1)$ (as it should 
be \cite{borgs}, a value close but smaller than that for a finite 
system, e.g., the one in Fig. \ref{hist-pt-st}).
After locating $\mu^{*}$ with accuracy, we can perform a rescaling 
of the data, finding a very good collapse by plotting
$\rho \times (\mu-\mu^*) \, V$ and 
$\chi_T/V \times (\mu-\mu^*) \, V$.
Such results confirm the amenability of the present procedure 
for estimating the transition points in discontinuous phase
transitions.

\section{Remarks and Conclusion}

In this paper we have considered an alternative simple protocol
to calculate very accurate free-energy weights for the ST 
(simulated tempering) method. 
It consists in estimating $f$ by means of the largest eigenvalue
of the transfer matrix \cite{sauerwein}, a quantity directly 
obtained from standard Monte Carlo simulations.
To illustrate the approach, we have addressed strong first-order 
phase transitions for different lattice models, namely, Blume-Capel, 
BEG and Bell-Lavis.
In such regime, distinct regions of the phase space can be separated 
by high entropic barriers, thus demanding efficient sampling 
procedures.
In all cases, our results have agreed quite well with available
precise calculations in the literature, with the advantage that
our ST has a simple implementation.
Next, two remarks are in order.

First, as already mentioned, an advantage of the present implementation
is the relative low computation cost in obtaining the refined weights
-- based on standard MC simulations.
It is not our purpose here a detailed benchmark comparison 
between different methods.
Nevertheless, without going into the merit of the (good) performance 
of other approaches such as Wang-Landau and parallel tempering (for the 
later, see, e.g., Ref. \cite{fiore-luz}), we observe the following.
Algorithmically speaking, in principle the ST may be faster since 
it does not demand:
either (i) to know the density of states as in the Wang-Landau  
(usually a hard task, e.g., in our examples involving $3^V$ 
configurations);
or (ii) to simultaneously simulate many replicas as in the PT 
(in the ST only one system realization is considered).

Second, by increasing too much the system size, the accepting 
probabilities for the temperature exchanges decrease, eventually 
leading to a poor estimation for the thermodynamic quantities 
(but for a possible way to assuage it, see the end of Section II).
In fact, the difficulty in simulating large systems is not a 
peculiarity of the ST.
It also may be the case in others methods like the PT and Wang-Landau. 
A finite-size analysis circumvent this problem, but in practice should 
use $L$'s up to a certain maximum value $L_m$, for which the considered 
method still works well.
The crucial point is whether $L_m$ allows a correct 
extrapolation to the thermodynamic limit.
In our many examples, for $L$ around 25 we already can predict this 
limit.
But other situations would require, say, $L=100$, which in principle 
could be calculated with our approach if we use a larger set of 
temperatures $\{T_n\}$, but with $\Delta T = (T_n - T_1)$ fixed.
Presently, such issue is under investigation and will the subject of a 
forthcoming publication.

In summary, the examples given show that the ST algorithm allied
to our straightforward way to calculate the weights is 
able to deal with free-energy barriers, common at the phase 
coexistence, and therefore can be very useful in characterizing
first-order phase transitions.

\section*{Acknowledgements}

We acknowledge researcher grants from CNPq.
Financial support is also provided by CNPq-Edital Universal, 
Funda\c c\~ao Arauc\'aria and Finep/CT-Infra.

\appendix

\section{The transfer matrix method to calculate $Z$}

Here we present a general overview on the transfer matrix 
method~\cite{sauerwein}, and how it can be used to calculate $Z$.

Consider a general Hamiltonian of a regular lattice, written as
\begin{equation}   
{\cal H}= \sum_{k=1}^K {\cal H}(S_k,S_{k+1}),   
\label{e14}  
\end{equation}  
for $S_k \equiv (\sigma_{1,k}, \sigma_{2,k}, \ldots, \sigma_{L,k})$
the state configuration of the $k$-th layer (each having $L$ 
sites).
Also, assume periodic boundary conditions, so that $S_{K+1} = S_1$. 

From the definition of the (grand-canonical) partition function $Z$, 
we have that the probability $P(S_{1},S_{2},...,S_{K})$ for the 
layer 1 to have the configuration $S_1$, the layer 2 the 
configuration $S_2$, and so on, is given by ($\beta = 1/(k_B T)$)
\begin{equation}    
P(S_{1},\ldots,S_{K}) = 
\frac{
\exp[- \beta {\cal H}(S_1, S_2)] \times \ldots \times
\exp[- \beta {\cal H}(S_{K-1}, S_K)] \times 
\exp[- \beta {\cal H}(S_K, S_1)]}{Z}.
\label{prob}
\end{equation}
We then can define the transfer matrix ${\cal T}$ such that
its elements ${\cal T}(S',S'')$ are equal to
$\exp[- \beta {\cal H}(S', S'')]$, for $S'$ and $S''$ being the
configurations of two successive neighbor layers.
Hence, the r.h.s. of Eq. (\ref{prob}) reads 
${\cal T}(S_{1},S_{2}) \ldots {\cal T}(S_{K},S_{1}) / Z$.

Next, since
\begin{equation} 
\sum_{S_1, \ldots, S_K} P(S_{1}, \ldots, S_{K}) = 1,
\end{equation}
it naturally follows that
\begin{equation}    
Z = \sum_{S_1, \ldots, S_K}   
{\cal T}(S_{1},S_{2}) \, {\cal T}(S_{2},S_{3}) \, \ldots \, 
{\cal T}(S_{K-1},S_{K}) \, {\cal T}(S_{K},S_{1}).   
\label{e15}  
\end{equation}  
But observe that 
$\sum_{S_k} {\cal T}(S_{k-1}, S_k) \, {\cal T}(S_{k}, S_{k+1}) =
{\cal T}^2(S_{k-1}, S_{k+1})$, with this last term denoting the element 
$(S_{k-1}, \, S_{k+1})$ of the matrix ${\cal T}^2$.
Thus, using such relation recursively in Eq. (\ref{e15}), one finds
\begin{equation}   
Z = \sum_{S_1} {\cal T}^K(S_1, S_1) = \mbox{Tr}[{\cal T}^K].   
\label{e16}  
\end{equation}   
If now we calculate all the eigenvalues $\{ \lambda^{(m)} \}$ of 
${\cal T}$, from basic linear algebra we get
\begin{equation}   
Z =  \sum_{m} (\lambda^{(m)})^K.  
\label{e20}    
\end{equation}
Finally, for $K$ large enough, the most important contribution
in Eq. (\ref{e20}) comes from the largest eigenvalue of ${\cal T}$,
$\lambda^{(0)}$, and thus we recover Eq. (\ref{z-lambda}) of 
Sec. II (which is exact in the thermodynamic limit of 
$K \rightarrow \infty$).

To derive an expression for $\lambda^{(0)}$, observe that the
marginal probabilities
\begin{equation}
P(S_1) = \sum_{S_2, \ldots, S_N} P(S_{1}, S_2, S_3, \ldots, S_{N}),
\qquad
P(S_1, S_2) = \sum_{S_3, \ldots, S_N} P(S_{1}, S_2, S_3, \ldots, S_{N}),
\end{equation}
are readily obtained from Eq. (\ref{prob}) and from proper products 
of the ${\cal T}$ matrix elements, or
($S_1 = S'$, $S_2 = S''$)
\begin{equation}   
P(S') = \frac{{\cal T}^{K}(S',S')}{Z}, \qquad
P(S',S'')= \frac{{\cal T}(S',S'') \, {\cal T}^{K-1}(S'',S')}{Z}.   
\end{equation}   
For $| m \rangle$ being the eigenvector associated to the eigenvalue 
$\lambda^{(m)}$ of ${\cal T}$, we have that the usual spectral
expansion of an operator,  
${\cal T} = \sum_m \lambda^{(m)} \, |m \rangle \langle m |$,
yields
\begin{equation}   
{\cal T}(S',S'') = \sum_m \, \lambda^{(m)} \, 
                    \phi^{(m)}(S') \, {\phi^{(m)}}^{*}(S''),  
\label{e19}   
\end{equation}   
for $\phi^{(m)}(S')$ an element of $| m \rangle$ in the representation
of the layer configurations $\{ S \}$.
Therefore   
\begin{equation}   
P(S') = \frac{1}{Z} \, \sum_{m} \, (\lambda^{(m)})^K \,
\phi^{(m)}(S') \, {\phi^{(m)}}^{*}(S'),  
\label{e21}      
\end{equation}   
and   
\begin{equation}   
P(S',S'')=\frac{1}{Z} \, {\cal T}(S',S'') \sum_{m} 
\, (\lambda^{(m)})^{K-1} \,
\phi^{(m)}(S')  \, {\phi^{(m)}}^{*}(S'').  
\label{e22}     
\end{equation}   
Again, considering $K$ large enough, Eqs. (\ref{e21}) and (\ref{e22})
can be approximated by  
\begin{equation}   
P(S') = \phi^{(0)}(S') \, {\phi^{(0)}}^{*}(S'), \qquad
P(S',S'') = \frac{1}{\lambda^{(0)}} \, {\cal T}(S',S'') \, 
\phi^{(0)}(S') \, {\phi^{(0)}}^{*}(S'').  
\label{e23}    
\end{equation}   
Setting $S' = S''$ in Eq. (\ref{e23}), we arrive at  
\begin{equation} 
P(S',S') =  
\sum_{S''} \delta_{s',s''} P(S',S'') = 
\frac{1}{\lambda^{(0)}} {\cal T}(S',S') P(S').  
\label{e25}    
\end{equation}   
Lastly, summing Eq. (\ref{e25}) over $S'$ and identifying the averages
(for ${\cal T}(S') \equiv {\cal T}(S',S')$)   
\begin{equation}  
\left< \delta_{S',S''} \right> = 
\sum_{S',S''} \delta_{s',s''} \, P(S',S''), \qquad
\left< {\cal T}(S') \right> = \sum_{S'}  {\cal T}(S') \, P(S'),  
\label{e26}   
\end{equation}  
we obtain Eq. (\ref{lambzero}).


\end{document}